\documentclass[aps,pra,twocolumn,showpacs]{revtex4}

\usepackage{amsmath,psfig} 
\usepackage{graphicx} 
 
\allowdisplaybreaks

\newcommand{\bx}{{\bf x} } 
 
\begin{document}

\title{Three Dimensional Simulation of Jet Formation in 
Collapsing Condensates} 
 
\author{Weizhu  Bao$^1$, D.~Jaksch$^2$, and P.A.~Markowich$^3$} 
 
\affiliation{ $^1$ Department of Computational Science, National 
University of Singapore, Singapore 117543.\\ 
$^2$Clarendon Laboratory, Department of Physics, University of 
Oxford, Oxford, United Kingdom.\\ 
$^3$Institute of Mathematics, University of Vienna Boltzmanngasse 
9, A-1090 Vienna, Austria.} 
\date{\today}

\begin{abstract} 
 
We numerically study the behavior of collapsing and exploding 
condensates using the parameters of the experiments by E.A.~Donley 
{\em et al.} [Nature, {\bf 412}, 295, (2001)]. Our studies are 
based on a full three-dimensional numerical solution of the 
Gross-Pitaevskii equation (GPE) including three body loss. We 
determine the three body loss rate from the number of remnant 
condensate atoms and collapse times and obtain only one possible 
value which fits with the experimental results. We then study the 
formation of jet atoms by interrupting the collapse and find very 
good agreement with the experiment. Furthermore we investigate the 
sensitivity of the jets to the initial conditions. According to 
our analysis the dynamics of the burst atoms is not described by 
the GPE with three body loss incorporated. 
\end{abstract}

 
\maketitle 
 
\section{Introduction} 
 
Most of the single particle properties of Bose-Einstein 
condensates (BEC) in dilute weakly interacting gases are well 
described by the Gross-Pitaevskii equation (GPE) (for reviews see 
\cite{Anglin, Dalfovo, Cornell}). The GPE is well suited for 
investigating static as well as dynamic properties of a BEC and 
also allows to investigate the stability of BECs with attractive 
interactions in a trapping potential. Inelastic processes which 
only become important for large particle densities are usually 
accounted for by adding cubic and quintic damping terms to the GPE 
which are believed to properly describe two- and three-particle 
loss, respectively. 
 
In early experiments on BEC with attractive interactions 
\cite{Hulet} the scattering length was fixed and limited the size 
of these condensates to a number $N_{\rm cr}$ \cite{Kagan} which 
for typical experimental parameters was on the order of $N_{\rm 
cr} \approx 1000$. In these experiments a collapse of the 
condensate was induced stochastically while growing the BEC out of 
a thermal cloud of atoms. In contrast, more recent experiments by 
Donley {\em et al.} \cite{Donley} allowed to deterministically 
induce collapses of the condensate which revealed a dramatic 
behavior for a ${}^{85}$Rb BEC. In these experiments the sign of 
the scattering length is changed from positive (repulsive 
interaction) to negative (attractive interaction) values by an 
external magnetic field. This sudden change in the sign of the 
interaction leads to a series of collapses and explosions, a 
dynamical behavior which provides a very good opportunity for 
testing the applicability of the GPE. 
 
In particular the nature of the atoms emitted in a burst during 
the collapse of the condensate is not very well understood at 
present \cite{Savage}. While some numerical studies \cite{Ueda, 
Santos} indicate that these atoms are produced coherently and can 
be described by the GPE, other numerical \cite{Adhikari} and 
analytical approaches \cite{Stoof, Holland, Javanainen, Burnett, 
calzetta} conclude that these atoms are not described by the GPE 
alone. 
 
The numerical studies arrive at different conclusions mostly due 
to the choice of the three particle loss rate near the Feshbach 
resonance. The work of Saito and Ueda \cite{Ueda} predicts a 
series of collapses and explosions during the experiment and 
correctly reproduces the burst atoms by choosing a three particle 
loss rate that is smaller than the one used by Adhikari 
\cite{Adhikari} where one big collapse determines most of the 
behavior of the condensate and the burst atoms are not reproduced 
correctly. Varying the three particle loss rate with the 
scattering length \cite{Santos} to match the burst energies allows 
to get good quantitative agreement with the experimental data. 
 
In the analytical approach by Duine and Stoof \cite{Stoof} elastic 
binary collisions in the BEC are suggested to cause the bursts and 
in \cite{calzetta} the dynamics of collective excitations driven 
by the collapsing condensate are investigated analytically and 
found to explain the overall features of collapsing and exploding 
condensates. The effect of a molecular state close to threshold 
near a Feshbach resonance on the interaction between the atoms is 
explored in \cite{Holland, Javanainen, Burnett}. In fact, such 
novel interaction mechanisms arising close to Feshbach resonances 
which are not contained in the GPE could yield the relatively 
large burst energies seen in the experiment \cite{Holland}. 
 
\begin{figure} 
\includegraphics[width=8.5cm]{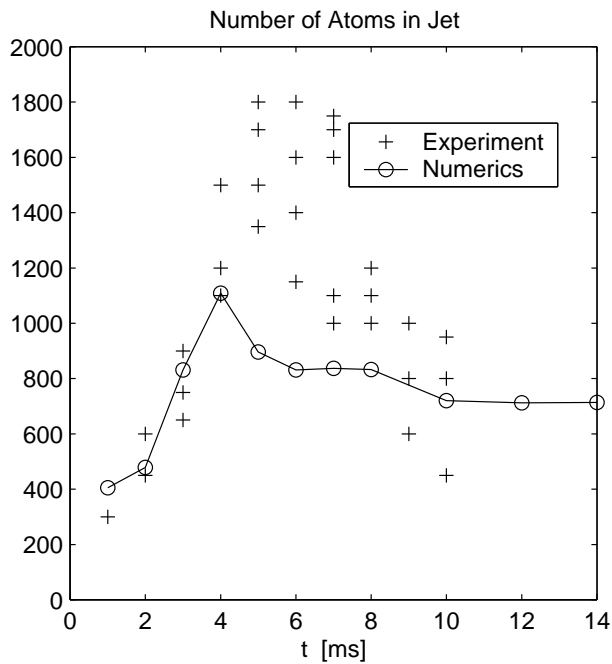} 
\caption{Number of atoms in the jets as a function of time 
$\tau_{\rm evolve}$ [ms] (labelled as $t$) for $a_{\rm init}=7a_0$, 
$N_0=16000$ and $a_{\rm collapse}=-30 a_0$. `+ +': Experimental 
data from \cite{Donley}. `--o--': Numerical results for parameters 
equivalent to those used in the experiment. \label{fig6}} 
\end{figure}

In this paper we study numerically the full three dimensional GPE 
with attractive interactions including three particle loss. We 
first choose the three-body loss rate $K_3^0$ to match the 
observed remnant condensate particle numbers and collapse times. 
We find only one single value for $K_3^0$ which fits the 
experimental data and subsequently use this value in our numerics. 
Then we investigate the time evolution of the collapsing 
condensate. We find jets of atoms whose particle number and 
sensitivity to initial conditions is in quantitative agreement 
with the experiment by Donley \rm {et al.} \cite{Donley}. The 
results of our comparison can be seen in Fig.~\ref{fig6} where we 
plot the number of jet atoms as a function of the time $\tau_{\rm 
evolve}$ at which the collapse of the condensate is interrupted. 
Also, the shape and energies of the jets agree with the 
experiment. We do not, however, find burst atoms with energies as 
measured in the experiment and thus conclude that they are not 
described by the physics contained in the GPE alone. 
 
For our numerical studies we use the time-splitting sine-spectral 
method (TSSP) recently introduced by Bao {\em et al.} for solving 
the damped GPE \cite{Bao3,BJ}. This method is explicit, 
unconditionally stable, time transversal invariant, and of 
spectral accuracy in space and second order accuracy in time. It 
thus yields reliable results even in the case of having sharply 
peaked wave functions during the collapse. 
 
The paper is organized as follows: In Sec.~\ref{nummeth} we will 
introduce the GPE including loss terms and present the numerical 
method we use to solve it. Then, in Sec.~\ref{results} we first 
adjust the three particle loss rate to match the experimental 
results for the number of remnant particles and the collapse time. 
This is followed by a detailed comparison of numerical and 
experimental results on the jet atoms and bursts of atoms. Finally 
in Sec.~\ref{concl} we draw some conclusions.

\section{Model and numerical method} 
\label{nummeth} 
 
In this section we will first introduce the GPE used for 
describing the dynamics of a BEC with varying interaction 
strength. Then we briefly present the numerical method for solving 
this three dimensional nonlinear partial differential equation. 
 
\subsection{Model} 
 
We consider a one component BEC with varying scattering length 
that is described by the GPE including a damping term that 
accounts for inelastic interactions 
\begin{equation} 
i\hbar \frac{\partial \psi(\bx,t)}{\partial t}=\left(-\frac{\hbar^2\nabla^2} 
{2 m}+ V(\bx) + U(t) |\psi|^{2} -i\; \frac{g(|\psi|^2)}{2} \right) 
\psi,  \label{sdge} 
\end{equation} 
for times $t>0$ with the initial condition \begin{equation} 
\psi(\bx,t=0)=\psi_0(\bx). \label{sdgi} \end{equation} Here 
$\psi(\bx,t)$ is the macroscopic wave function of the condensate, 
and $\bx=(x,y,z)^T$ is the spatial coordinate. We assume the 
trapping potential to be harmonic $V({\bf x})=m\left(\omega_x^2 
x^2 +\omega_y^2 y^2+\omega_z^2 z^2 \right)/2$ with $\omega_x$, 
$\omega_y$, $\omega_z$ the trapping frequencies in $x$, $y$, and 
$z$ direction respectively, and $m$ the mass of the atoms. 
 
\begin{figure} 
\includegraphics[width=8.5cm]{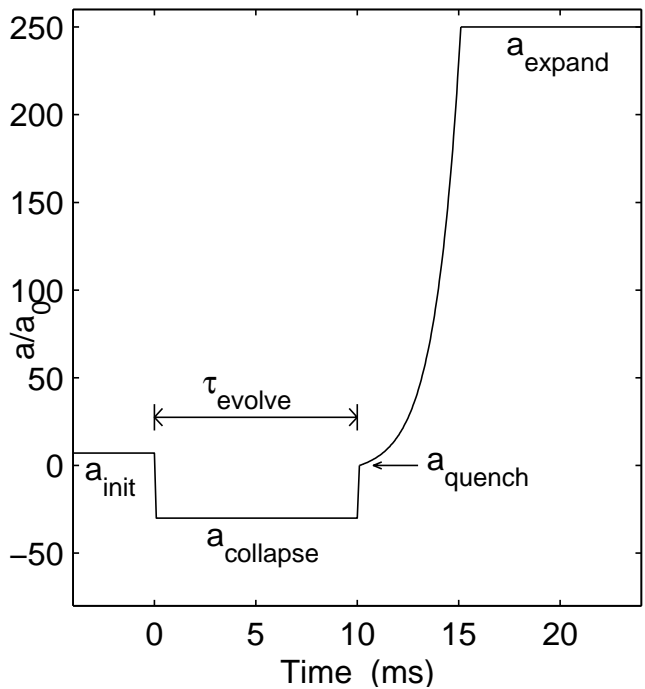} 
\caption{Example of a time dependence of the scattering length 
$a_s$ during the experiment \cite{Donley}. A condensate with 
ground state wave function $\psi_0(\bx)$ is prepared at 
$a_s=a_{\rm init}$. Then the collapse is induced by going to a 
negative scattering length for a time $\tau_{\rm evolve}$ and 
finally the time evolution of the remaining particles is studied 
at large positive values of $a_s$. \label{fig201}} 
\end{figure}

The macroscopic wave function at time  $t=0$, i.e. the initial 
data $\psi_0$, is normalized 
\begin{equation} 
\int_{{\bf R}^3}\; |\psi(\bx,0)|^2 \; d^3{\bf x}=\int_{{\bf R}^3}\; 
|\psi_0(\bx)|^2 \; d^3{\bf x} =N_0, 
\end{equation} 
where $N_0$ is the total number of condensate particles at time 
$t=0$. The two-body interaction between the atoms is given by 
$U(t)=4 \pi \hbar^2 a_s(t)/m$ with $a_s(t)$ the $s$-wave 
scattering length.  In the experiment the time dependence of 
$a_s(t)$ (cf. Fig. \ref{fig201}) \cite{Donley} is realized by 
changing the magnitude of an external magnetic field near a 
Feshbach resonance of the $^{85}$Rb atoms. 
 
In the case of a positive scattering length $a_s$ the GPE 
(\ref{sdge}) with $g\equiv0$ has a stable ground state solution 
$\psi_{\rm gs}(\bx)$ while if the sign of $a_s(t)$ is changed from 
positive to negative the GPE becomes focusing and does not have a 
stable ground state solution anymore, i.e., the condensate may 
collapse. As in the experiment we will start our simulations at 
positive scattering lengths $a_s$ and for most of the calculations 
assume the BEC initially to be in its ground state \cite{BD,Bao} 
corresponding to the initial condition Eq.~(\ref{sdgi}) with 
$\psi_0(\bx)=\psi_{\rm gs}(\bx)$. 
 
\begin{figure} 
\includegraphics[width=8.5cm]{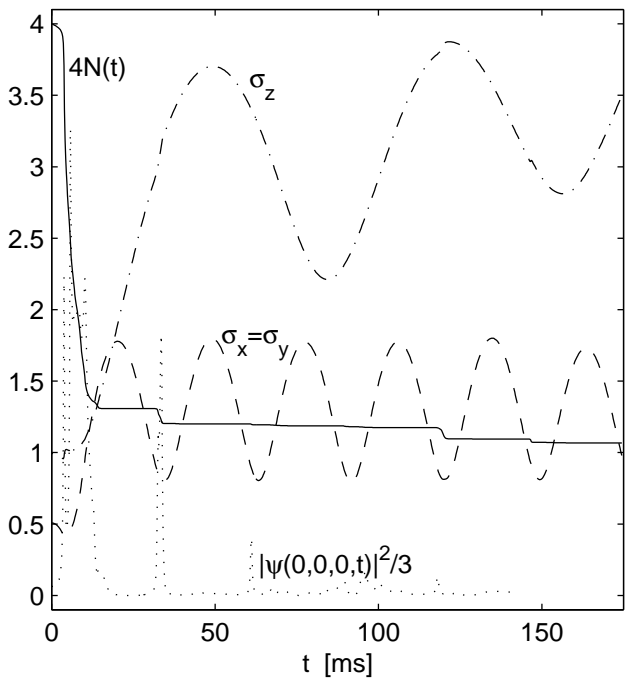} 
\caption{Normalization, condensate widths  and central density 
$|\psi(0,0,0,t)|^2$ as functions of time 
$\tau_{\rm evolve}$ [ms] (labelled as $t$) for $a_{\rm init}=7a_0$, 
$a_{\rm collapse}=-30a_0$ and $N_0=16000$. \label{fig2a}} 
\end{figure} 
 
Loss from the condensate is described by the term 
$g(\rho)=\hbar\left(K_2(t)\rho +K_3(t)\rho^2\right)$ where 
$K_2(t)$ is due to two-body dipolar loss and $K_3(t)$ accounts for 
the three-body recombination inelastic processes. We assume the 
effects of $K_2(t)$ to be negligible and set $K_3(t)$ equal to 
$K_3^0$ when $a_s(t) = a_{\rm collapse}$, and $0$ otherwise 
\cite{Donley,Ueda,Stoof}. Furthermore we assume $K_3^0$ to be 
proportional to $a^2_{\rm collapse}$ \cite{Ueda,Adhikari}. Under 
this assumption we choose $K_3^0$ to match the number of remnant 
atoms in the condensate after the collapse and the collapse times 
observed in the experiments. We also compare its value with 
experimental data in \cite{Roberts}. 
 
\subsection{Dimensionless GPE} 
 
In our numerical computations, we simulate and present numerical 
results based on the following dimensionless GPE. We introduce 
$\tilde {t}= \omega_z t$, $\tilde{\bx}=\bx/l_0$, 
$\tilde{\psi}(\tilde{\bx},\tilde{t})=l_0^{3/2}\psi(\bx,t)/\sqrt{N_0}$, 
%$\tilde{a}_s(\tilde{t})=a_s(t)$, 
$\tilde{\psi}_0(\tilde{\bx}) =l_0^{3/2}\psi_0(\bx)/\sqrt{N_0}$, 
with $l_0=\sqrt{\hbar/m\omega_z}$ the size of the harmonic 
oscillator ground state. 
%In fact, here we choose $1/\omega_z$ and $l_0$ as 
the dimensionless time and length units respectively. 
Substituting into Eq.~(\ref{sdge}), multiplying both sides by 
$l_0^{3/2}/\hbar \omega_z \sqrt{N_0}$, and removing all \~{ }, we 
obtain the following dimensionless GPE 
\begin{equation} 
i\;\frac{\partial \psi(\bx,t)}{\partial t}=\left(-\frac{\nabla^2} 
{2}+ V(\bx) + \beta(t) |\psi|^{2} 
-i\; \frac{g(|\psi|^2)}{2} \right)\psi, 
\label{sdgedl} 
\end{equation} 
with the initial condition 
\begin{equation} 
\psi(\bx,t=0)=\psi_0(\bx); \label{sdgidl} 
\end{equation} 
where $\beta(t)=4\pi a_s(t) N_0/l_0$, $g(\rho) = \delta_1(t) 
\beta_c N_0 \rho +\delta_2(t) \beta_c^2 N_0^2\rho^2$, 
$\delta_1(t)$ and $\delta_2(t)$ equal to $\delta_1^0$ and 
$\delta_2^0$ respectively when $a_s(t) = a_{\rm collapse}$, and 
$0$ otherwise, $V(\bx) = \frac{1}{2}\left(\gamma_x^2 x^2 
+\gamma_y^2 y^2 +z^2\right)$ with $\gamma_x=\omega_x/\omega_z$ and 
$\gamma_y =\omega_y/\omega_z$, $\beta_c = 4\pi |a_{\rm 
collapse}|/l_0$, $\delta_1^0 = K_2^0 / \left(l_0^3 \omega_z 
\beta_c\right)$, $\delta_2^0 = K_3^0 / \left(l_0^6 \omega_z 
\beta_c^2\right)$. For $a_s(t) = a_{\rm collapse}<0$ we therefore 
have $\beta(t) = -\beta_c N_0$. The macroscopic wave function 
$\psi$ in the dimensionless GPE (\ref{sdgedl}) is now normalized 
to $1$ at time $t=0$, i.e., 
\begin{equation} 
\int_{{\bf R}^3}\; |\psi(\bx,0)|^2 \; d^3{\bf x}=\int_{{\bf R}^3}\; 
|\psi_0(\bx)|^2 \; d^3{\bf x} =1. 
\end{equation} 
We assume that $\delta_1^0$ and $\delta_2^0$ are independent of 
$N_0$ and $a_{\rm collapse}$ \cite{Ueda,Adhikari}. 
 
The fraction of particles $N_{\Omega}(t)$ in a volume $\Omega$ is 
given by 
\begin{equation} \label{Ntom} 
N_\Omega(t)=\int_{\Omega}\; |\psi({\bf x}, t)|^2\; d^3{\bf x}, 
\end{equation} 
and therefore the number of condensate particles in a volume 
$\Omega$ is $N_0\; N_\Omega(t)$. Due to the loss term 
$g(|\psi|^2)$ the total number of particles $N_{\rm total}(t) = 
N_0\; N(t) =N_0\; N_{{\bf R}^3}(t)$ is time dependent, decaying as 
\begin{equation}\label{Nt} 
\dot N_{\rm total}(t)= -N_0 \int_{{\bf R}^3}\; g(|\psi({\bf x}, 
t)|^2)|\psi({\bf x}, t)|^2\; d^3{\bf x} \le 0. 
\end{equation} 
We do not further care for the particles lost from the BEC by 
inelastic collisions and concentrate on the dynamics of the 
remaining condensate particles. However, particles incoherently 
lost from the condensate might still be observed in an experiment. 
We also introduce the mean width of the condensate in directions 
$x,y,z$ (cf. Fig. \ref{fig2a}) which is computed from the wave 
function as 
\begin{equation}\label{rms} 
\sigma_\alpha^2(t)=\int_{{\bf R}^3}\; \alpha^2\; 
|\psi({\bf x}, t)|^2\; d^3{\bf x}/N(t), \quad \alpha=x,y,z. 
\end{equation} 
 
\subsection{Numerical method} 
 
We use the time splitting sine-spectral method (TSSP) which is 
described in detail in \cite{Bao3,BJ} for solving the damped GPE. 
This method for numerically solving Eq.~(\ref{sdgedl}) is based on 
a time-splitting, where at every time step one solves 
\begin{equation} i\; \frac{\partial \psi(\bx,t)}{\partial t} 
=-\frac{{\nabla}^2}{2}\psi,  \label{sdgef} \end{equation} 
followed by evolving \begin{equation} 
i\; \frac{\partial \psi(\bx,t)}{\partial t}=\left( V(\bx) + 
\beta(t) |\psi|^{2} -i\; \frac{g(|\psi|^2)}{2} \right) \psi.  \label{sdges} 
\end{equation} 
The linear Schr\"{o}dinger equation without external potential 
Eq.~(\ref{sdgef}) can be discretized in space by the sine-spectral 
method and then sovled in time {\em exactly} when homogeneous 
Dirichlet boundary conditions are applied \cite{BJ}. For each 
fixed ${\bf x}$, the ordinary differential equation (ODE) 
Eq.~(\ref{sdges}) can be integrated {\em exactly}, too \cite{BJ}. 
In fact, the TSSP for the GPE combines the advantages of the 
spectral method which yields spectral order accuracy in space and 
can generate simple and explicit numerical formulae for partial 
differential equations (PDEs) with constant coefficients when a 
proper orthogonal basis is chosen, and the split-step method which 
can decouple nonlinear PDEs, e.g. the GPE, into linear PDEs with 
constant coefficients and a nonlinear ODE which can usually be 
solved analytically. The merit of the TSSP for solving the GPE is 
that it is explicit, unconditionally stable, time reversible and 
time transverse invariant when the GPE is, of spectral order 
accuracy in space and of second order accuracy in time, conserves 
the total number of particles for the GPE (\ref{sdge}) without 
damping, i.e.~for $g\equiv0$, and exactly reproduces the decay 
rate of the total number of particles for the GPE (\ref{sdge}) 
with a linear damping term, i.e. $g\equiv \alpha_0>0$. 
 
It is well known that the three-dimensional cubic nonlinear Schr\"{o}dinger
equation (NLS)  (i.e. GPE without external potential) with sufficiently 
large attractive two-body interaction leads to finite-time 
collapse of the spatial condensate density \cite{Fibich,FibichP,Sulem}, 
i.e. $|\psi|^2$ 
becomes a delta distribution for some finite value of time, and 
afterwards the solution ceases to exist. This collapse mechanism 
is halted by the three-body recombination term, which starts to 
kick in when the density becomes too large locally and then acts 
to instantaneously reduce the density in an explosion-like 
process \cite{Fibich,FibichP}. We remark that the mathematical analysis of the 
GPE with loss terms Eq. (\ref{sdgedl}) is not well developed yet, 
a systematic study has just 
begun right now. 
 
\section{Results} 
\label{results} 
 
Before presenting the results of our numerical studies we adjust 
the three particle loss rate by fitting the numerical results to 
the experimental data for the remnant number of condensate atoms 
and the time of the collapse. Then we study the formation of jet 
atoms and study their sensitivity to initial conditions. Finally 
we investigate bursts of atoms and present results of the 
simulations that do not agree with the experiment. 
 
\subsection{Experimental parameters and three particle loss rate} 
\label{res1} 
 
To determine the three particle loss rate we use the experimental 
results for a collapsing condensate of ${}^{85}$Rb particles 
measured in \cite{Donley} where a condensate of $^{85}$Rb, with a 
particle mass $m=1.406\times 10^{-25}$kg, trapping frequencies 
$\omega_x=\omega_y=2\pi \times 17.5/$s, and $\omega_z=2\pi \times 
6.8/$s, i.e.~a cylindrically symmetric geometry were used. 
%This implies $l_0=\sqrt{\hbar/m\omega_z}=0.4181\times 10^{-5}$[m]. 
In the experiment at time $t=0$ a BEC with $N_{\rm 
total}(0)=N_0=N_0\; N(0)=16000$ atoms and a scattering length of 
$a_s(0)=7a_0$ (where $a_0$ is the Bohr radius) is prepared. 
Afterwards, by changing the external magnetic field, the 
scattering length is linearly ramped down to $a_s=a_{\rm 
collapse}=-30a_0$ (which corresponds to $\beta_c=4.7717\times 
10^{-3}$) in time $t=0.1$ms and held there for a time $\tau_{\rm 
evolve}$ as schematically shown in Fig.~\ref{fig201}. This process 
leads to a strongly attractive and unstable condensate which 
undergoes a sequence of collapses and explosions before finally a 
remnant condensate with particle number $N^0_{\rm remnant}$ is 
left. The remnant condensate is then measured at a positive 
scattering length (cf.~Fig.~\ref{fig201}). 
 
\begin{figure} 
\includegraphics[width=8.5cm]{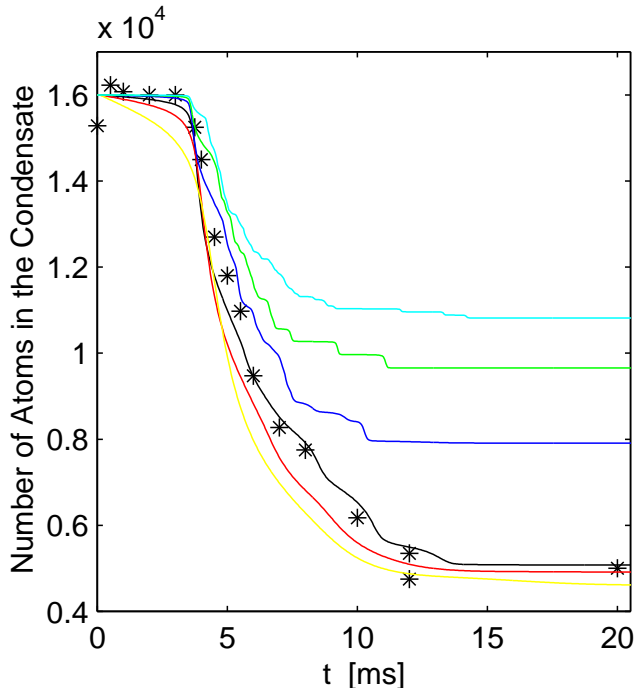} 
\caption{Number of remaining atoms after collapsing a $^{85}$Rb 
condensate of $N_0=16000$ atoms with different values for the 
three-body loss rate $\delta_2^0$. Collapse is achieved by ramping 
the scattering length linearly from $a_{\rm init}=7a_0$ to $a_{\rm 
collapse}=-30a_0$ in $0.1$ [ms] as a function of time 
$\tau_{\rm evolve}$ [ms] (labelled as $t$). The `*' 
are taken from the experiment \cite{Donley}. Curves are displayed 
in the order of decreasing $N_{\rm remnant}^0$ for:  Cyan: 
$\delta_2^0=0.00005$ (i.e. $K_3^0=2.598\times 10^{-28}$ 
[cm$^6$/s]); Green: $\delta_2^0=0.00016$ (i.e. $K_3^0=8.315\times 
10^{-28}$ [cm$^6$/s]); Blue: $\delta_2^0=0.0004$ (i.e. 
$K_3^0=2.079\times 10^{-27}$ [cm$^6$/s]); Black: 
$\delta_2^0=0.0013$ (i.e. $K_3^0=6.756\times 10^{-27}$ 
[cm$^6$/s]); Red: $\delta_2^0=0.003$ (i.e. $K_3^0=1.559\times 
10^{-26}$ [cm$^6$/s]); Yellow: $\delta_2^0=0.008$ (i.e. 
$K_3^0=4.157\times 10^{-26}$ [cm$^6$/s]). \label{fig202}} 
\end{figure}

We use the same time dependence for the scattering length $a_s$ in 
our numerics to find the quintic damping term $\delta_2^0$. 
Assuming cubic damping to be negligible we set $\delta_1^0=0$ and 
fit the parameter of the dominant quintic damping term 
$\delta_2^0$ according to the experimental results for $a_{\rm 
collapse}=-30 a_0$, i.e.~the number of remnant atoms in the 
condensate after the collapse and the collapse times observed in 
the experiments. We simulate the experiment and adjust the three 
particle loss rate to numerically obtain a number of remnant atoms 
that agrees with the experiment. Fig. \ref{fig202} shows the 
number of atoms in the condensate for different values of 
$\delta_2^0$. From this figure, we find a unique solution for the 
loss rate $\delta_2^0$ given by $\delta_2^0=1.3\times 10^{-3}$ 
which corresponds to a three particle loss rate of 
$K_3^0=\delta_2^0 l_0^6 \omega_z \beta_c^2=6.756\times 10^{-27}$ 
[cm$^6$/s]. This value for three particle loss is in agreement 
with the measured values shown in Fig.~2c of Ref.~\cite{Roberts} 
and we use it for all of the following computations. We will check 
agreement of the numerics with the experimental results for the 
remnant condensate particle number and the collapse times. 
 
\subsubsection{Remnant condensate particles} 
 
In Fig.~\ref{fig1} we show the comparison between the experimental 
and our numerical results for the number of remnant condensate 
particles $N_{\rm remnant}$ as a function of the time $\tau_{\rm 
evolve}$. The results are in quantitative agreement as can be seen 
from Fig.~\ref{fig1}. We fit our results for $N_{\rm remnant}$ to 
a smooth function of the form $N_{\rm remnant}(\tau_{\rm evolve})= 
N_{\rm remnant}^0+(N_0-N^0_{\rm remnant})\cdot \exp((t_{\rm 
collapse}- \tau_{\rm evolve})/t_{\rm decay})$ where $N^0_{\rm 
remnant}$ gives the number of condensate particles for $\tau_{\rm 
evolve} \rightarrow \infty$, $t_{\rm collapse}$ gives the time at 
which the condensate starts to collapse, and $t_{\rm decay}$ is 
the decay time constant that determines the loss of particles 
during the collapse. The values found in the experiment were 
$t_{\rm collapse}= 3.8, 8.6$ [ms] and $N^0_{\rm remnant}=5000, 
7000$ for $a_{\rm collapse}=-30a_0, -6.7a_0$, respectively which 
is in agreement with our numerical simulation where we get $t_{\rm 
collapse}= 3.6, 9.35$ [ms] and $N_{\rm remnant}^0=5075, 6970$ for 
$a_{\rm collapse}=-30a_0, -6.7a_0$, respectively. We find decay 
times $t_{\rm decay}=2.8, 2.8$ [ms] (cf.~Fig.~\ref{fig1}) for 
$a_{\rm collapse}=-30a_0, -6.7a_0$, respectively, which also 
agrees with the experiment. Furthermore, from our numerical 
simulation for $a_{\rm collapse}=-250a_0$ we find $N^0_{\rm 
remnant}\approx 1660$, $t_{\rm decay}=1.2$ [ms] and $t_{\rm 
collapse}=1.1$ [ms]. 
 
\subsubsection{Collapse time} 
 
Using the same value for $\delta_2^0$ we can also confirm the 
experimentally observed change in the time at which the condensate 
collapses as a function of the density of the initial condensate. 
In Fig.~\ref{fig2} we show the number of condensate particles for 
$N_0=6000$, $a_{\rm collapse}=-15a_0$ and two different initial 
condensate densities with $a_{\rm init}=0$ and  $a_{\rm init}=89a_0$. We find 
numerically $t_{\rm collapse}=5.7,\; 16.2$ [ms] for $a_{\rm 
init}=0,\; 89a_0$ respectively, which is in excellent agreement 
with the experiment \cite{Donley}. 
 
\subsubsection{Nature of the collapse} 
 
Our simulations reveal a series of collapses and explosions 
similarly to the experiment. In Fig.~\ref{fig2a} we plot the 
particle density in the center of the trap as well as the widths 
of the condensate wave function in the different directions 
$x,y,z$. The fraction of condensate particle number $N(t)$ is also shown. The 
first collapse is marked by a sharp increase in the particle 
density in the center of the trap. During this first collapse a 
large amount of the particles is lost from the condensate. For the 
parameters chosen in Fig.~\ref{fig2a} subsequent collapses are by 
far less important than the first one as they have only a minor 
effect on the particle numbers. Also the peak density of these 
collapses is much smaller than for the first collapse and the 
widths of the condensate wave function are hardly affected. The 
times at which subsequent collapses happen are determined by the 
stiffer oscillation frequencies in the harmonic trap (see 
Fig.~\ref{fig2a}), i.e.~as soon as those particles emitted during 
a collapse return to the $z$-axis the particle density in the 
center increases and the next collapse happens. 
 
The surface plots in Fig.~\ref{fig3} give a more detailed view of 
the evolution of the condensate density during the collapse. First 
the condensate contracts in the center of the trap and the 
particle density increases (Fig.~\ref{fig3}a,b). During the first 
collapse a number of sharp peaks forms in the vicinity of the trap 
center (Fig.~\ref{fig3}c,d) which subsequently, towards the end of 
the first collapse, spread out due to their kinetic energy 
(Fig.~\ref{fig3}e,f). As we will show later these peaks are not of 
sufficient kinetic energy to produce the bursts of atoms seen in 
the experiment but interference effects between them \cite{Ueda} 
lead to the formation of the jets. 
 
\begin{figure} 
\includegraphics[width=8.5cm]{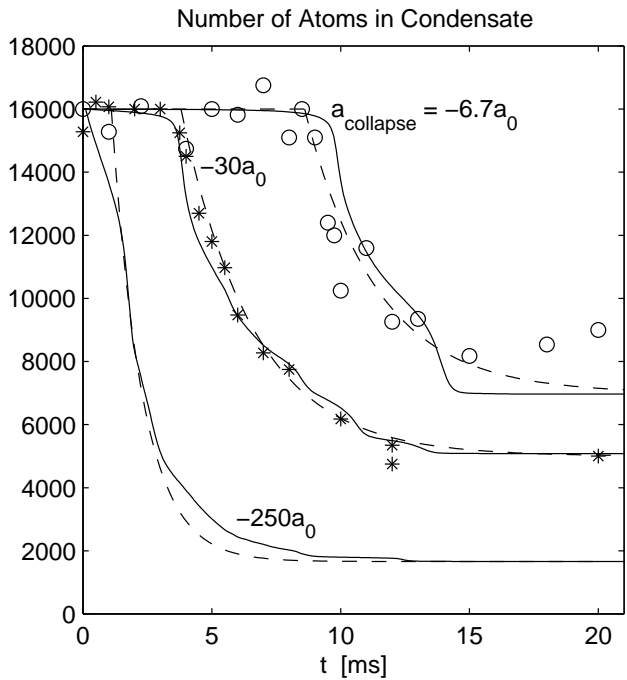} 
\caption{Number of remaining atoms after collapsing a $^{85}$Rb 
condensate of $N_0=16000$ atoms. Collapse is achieved by ramping 
the scattering length linearly from $a_{\rm init}=7a_0$ to $a_{\rm 
collapse}=-6.7a_0, -30a_0$ and $-250a_0$ in $0.1$ [ms] as a 
function of time $\tau_{\rm evolve}$ [ms] (labelled as $t$). 
The `*' and `o' are taken from the 
experiment \cite{Donley}, the solid curves are our numerical 
solutions and the dashed curves are fitted to the experimental 
points: $N_{\rm total}(t)=N_{\rm remnant}(\tau_{\rm evolve})
= N_{\rm remnant}^0+(N_0-N_{\rm 
remnant}^0)\cdot \exp((t_{\rm collapse}- \tau_{\rm evolve})
/t_{\rm decay})$ with 
$N_{\rm remnant}^0=7000,\; 5000,\; 1660$; $t_{\rm collapse}=8.6,\; 
3.8,\; 1.1$ [ms] and $t_{\rm decay} =2.8,\; 2.8,\; 1.2$ [ms] for 
$a_{\rm collapse}=-6.7a_0,\; -30a_0,\; -250a_0$, respectively. 
\label{fig1}} 
\end{figure} 
 
\begin{figure} 
\includegraphics[width=8.5cm]{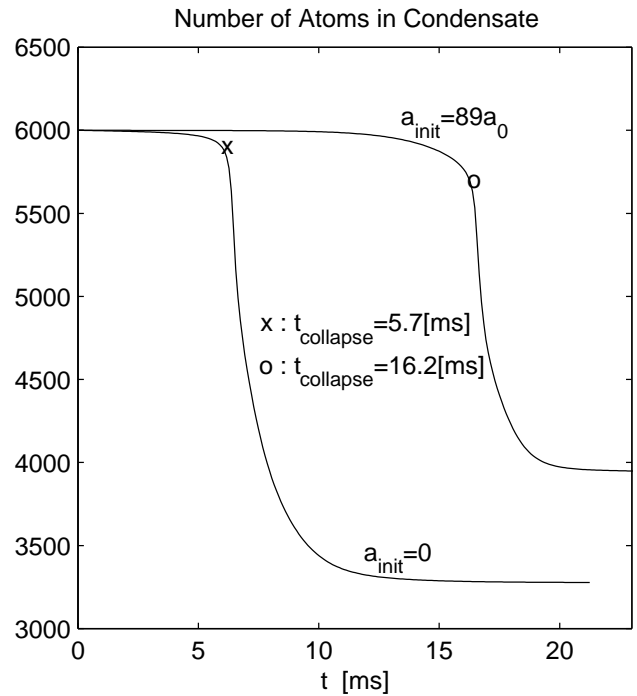} 
\caption{Number of remnant atoms in the $^{85}$Rb condensate for 
$N_0=6000$ after ramping the scattering length linearly from 
$a_{\rm init}=0$ and  $a_{\rm init}=89a_0$ at 
$t=0$ to $a_{\rm collapse}=-15a_0$ at 
$t=0.1$ [ms] as a function of time $\tau_{\rm evolve}$ [ms] 
(labelled as $t$). \label{fig2}} 
\end{figure} 
 
Finally we note that our three dimensional simulations allow this 
degree of agreement with the experiment only for the three 
particle loss rate $\delta_2^0$ chosen above. Simulations using 
smaller \cite{Ueda} or larger \cite{Adhikari} three particle loss 
rates do not yield numerical results that agree with the 
experimental data. 
 
\subsection{Jet formation} 
 
Next we are interested in the formation of jet atoms as observed 
in the experiment and thus simulate the following sequence for 
$a_s(t)$ (cf.~Fig.~\ref{fig201}): at $t=0$ the scattering length 
$a_s$ is ramped linearly from $a_{\rm init}$ to $a_{\rm collapse}$ 
in $0.1$ [ms], then kept constant for a time $\tau_{\rm evolve}$ 
(during this time period we apply the damping term in our 
numerics). Then the collapse is interrupted by switching $a_s$ 
linearly back from $a_{\rm collapse}$ to $a_{\rm quench}=0$ in 
$0.1$ [ms], followed by changing $a_s$ exponentially from $a_{\rm 
quench}$ to $a_{\rm expand}$ in $5$ [ms]. Then it is kept constant 
at $a_s=a_{\rm expand}$. The resulting condensate density is shown 
in Fig.~\ref{fig4} where we can see the emergence of jet atoms and 
their dynamics. 
 
For $\tau_{\rm evolve}< t_{\rm collapse}$, i.e.~before the 
condensate starts to collapse the outer region of the condensate 
is already affected by the negative scattering length and begins 
to expel some particles. As soon as the collapse has started this 
effect becomes more vigorous and condensate particles are ejected 
from the core of the condensate dominantly in the $xy$ plane 
forming jets. As the collapse continues, i.e.~$\tau_{\rm evolve}$ 
increases, the number of particles in the jet first becomes larger 
and then towards the end of the first collapse (cf.~also 
Fig.~\ref{fig1} and Fig.~\ref{fig2a} for the duration of the first 
collapse) decreases again. Finally, when the collapse is allowed 
to complete no jets can be seen anymore. The numerical results for 
the number of jet atoms as a function of $\tau_{\rm evolve}$ can 
be seen in Fig.~\ref{fig6}. We count the number of atoms in the 
jets from the atom density $|\psi|^2$ over a jet domain defined by 
$[-5 ,-0.75 ]^2\times[-0.75 ,0.75]\cup [0.75 ,5 
]^2\times[-0.75,0.75]$, and acquire the image $5.2$ [ms] after 
applying $a_{\rm quench}$) \cite{footnote}. These numerical 
results agree with the experimental data very well and also the 
jet pictures in Fig.~\ref{fig4} which give an impression on the 
shape of the jets agree well with those observed experimentally. 
 
\begin{figure} 
\includegraphics[width=8.5cm]{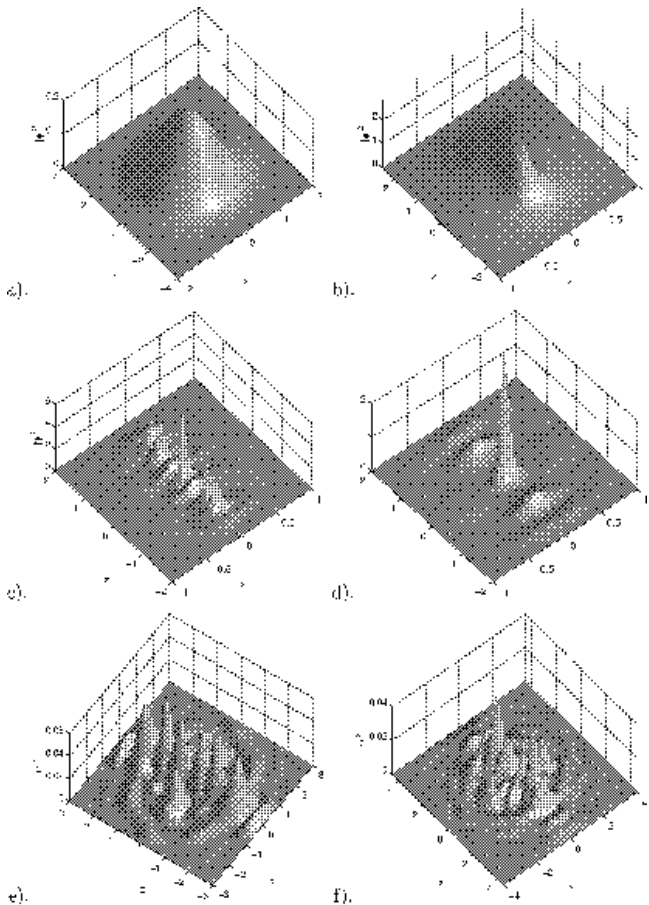} 
\caption{Surface plot of the density function $|\psi|^2$ in BEC 
with $a_{\rm init}=7a_0$, $a_{\rm collapse}=-30a_0$ and 
$N_0=16000$. At times  $\tau_{\rm evolve}$ 
(in [ms])\quad  a). $0$, b). $3.51$, 
c). $10.53$, d). $14.04$, e). $17.55$, f). $21.06$. \label{fig3}} 
\end{figure}

\begin{figure}[t] 
\includegraphics[width=8.5cm]{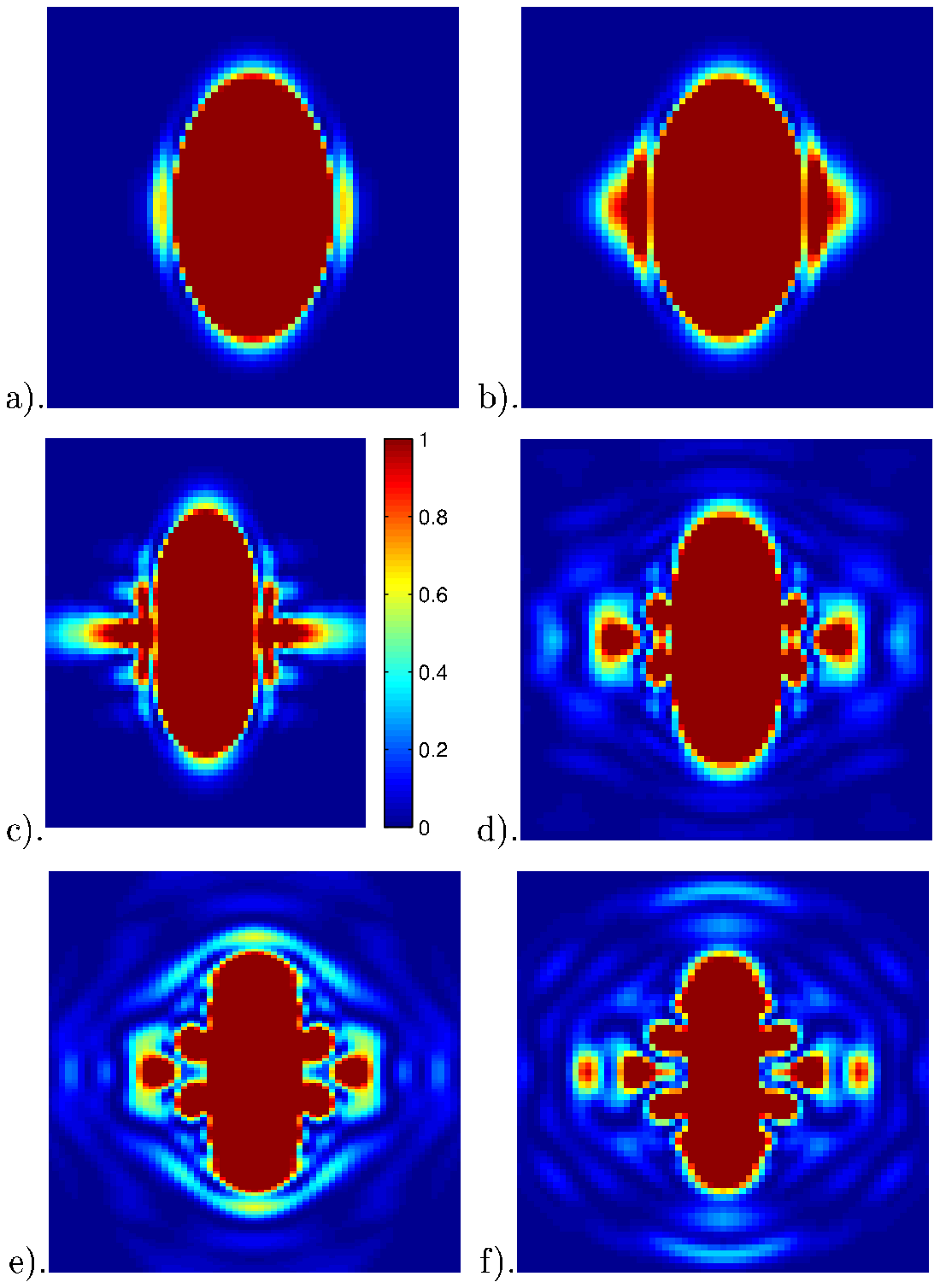} 
\caption{Jet images (i.e. image of $|\psi(x,0,z,t)|^2$) for a 
series of $\tau_{\rm evolve}$ 
values for $a_{\rm init}=7a_0$, $N_0=16000$ and 
$a_{\rm collapse}=-30 a_0$. The evolution times $\tau_{\rm evolve}$ 
 were $2$, $3$, 
$4$, $6$, $8$ and $10$ [ms] (from a) to f)). $a_{\rm quench}=0$ 
and $a_{\rm expand}=250 a_0$. The time from the application of 
$a_{\rm quench}$ until the acquisition of the images was equal to 
$5.2$ [ms]. \label{fig4}} 
\end{figure}

\begin{figure}[t] 
\includegraphics[width=8.5cm]{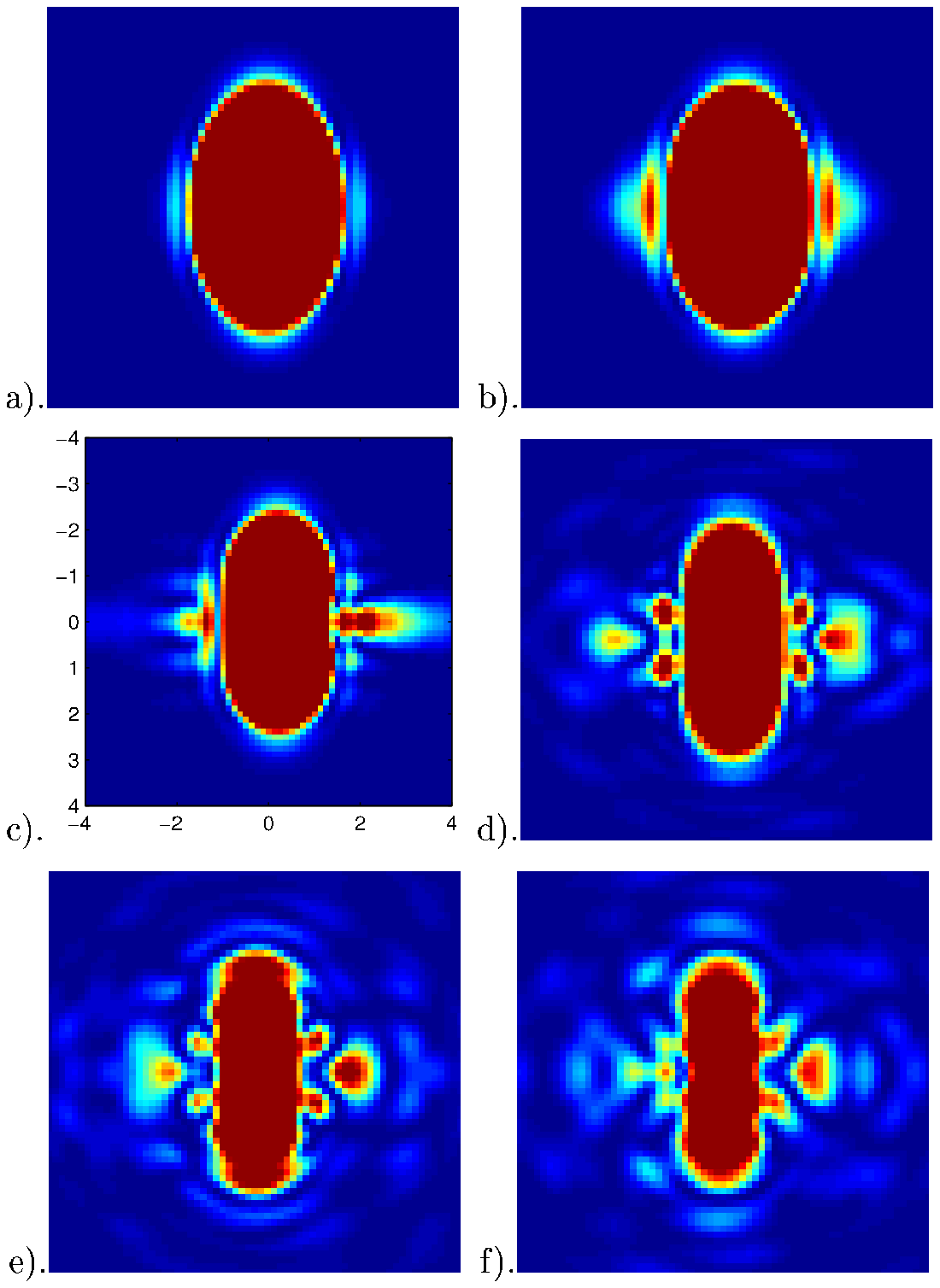} 
\caption{Jet images for time $\tau_{\rm evolve}$ 
 at $2$, $3$, $4$, $6$, $8$ and 
$10$ [ms] (from a) to f)) when we shift the center of the initial 
data from the origin to $(0.375,0,0)$. All other parameters as in 
Fig.~\ref{fig6}. \label{fig5}} 
\end{figure}

A large variance in the number of jet particles was found in the 
experiment. A possible reason for this could be that the initial 
condensate was not prepared exactly in the ground state. Then, if 
the jets are formed by interferences of particles emitted from 
different point like peaks in the atomic density \cite{Ueda} as 
shown in Fig.~\ref{fig3} slight deviations in the initial 
condition from the ground state wave function may have a big 
influence on the numbers of particles in the jets. We have tested 
this assumption by introducing a small offsets of the initial 
condensate wave function from the center of the trap and indeed 
found large variations in the jet particle number. A typical 
example is given in Fig.~\ref{fig5} where an asymmetric jet of 
atoms can be seen. For small times $\tau_{\rm evolve}$ the 
behavior of the condensate is very similar to the case of a 
centered BEC shown in Fig.~\ref{fig4}. This is in accordance with 
the small variance in the particle number for $\tau_{\rm evolve}< 
4$ms observed experimentally. For larger times $\tau_{\rm evolve}$ 
the behavior of the condensate with an initial offset is 
significantly different from the centered case; an asymmetric jet 
is formed in Fig.~\ref{fig5} and we find large changes in the 
number of jet particles for small offsets in qualitative agreement 
with the experimentally observed values. The offset chosen in 
Fig.~\ref{fig5} corresponds to an initial potential energy of the 
condensate of $E_p \approx 0.47 \hbar \omega_z$ and thus to a 
temperature of $T \approx 0.15$nK. Therefore it seems 
plausible that even small temperatures - although they might not 
lead to an offset of the condensate as chosen in our numerical 
example - and/or uncertainties in the initial wave function can 
influence the jet formation and lead to the large variance seen in 
the experiment. 
 
\subsection{Bursts of atoms} 
 
Finally we also want to study the bursts of atoms observed in the 
experiments \cite{Donley}. The bursts are particles that are 
emitted from the condensate at relatively high energies when the 
collapse is allowed to complete. For finding the burst particles 
we compute $\phi_z(z,t)$ and $\phi_{xy}(x,y,t)$ as the axially and 
radially averaged density cross-sections, respectively, as 
follows: 
\begin{eqnarray} &&\phi_z(z,t)=\int_{-\infty}^\infty 
\int_{-\infty}^\infty 
|\psi(x,y,z,t)|^2\; dxdy,\\ 
&&\phi_{xy}(x,y,t) = \int_{-\infty}^\infty |\psi(x,y,z,t)|^2\;dz. 
\end{eqnarray} 
 
\begin{figure} 
a).\includegraphics{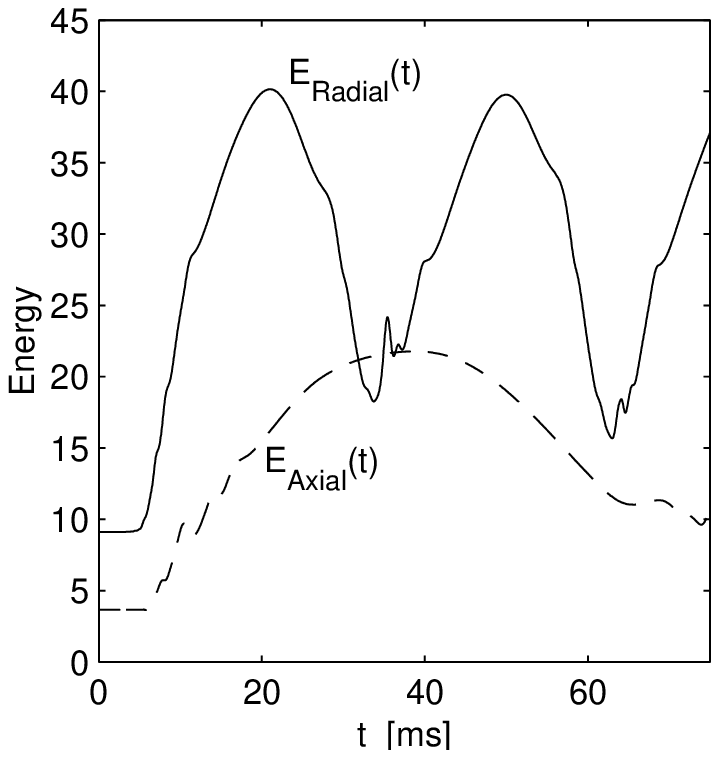} 
b).\includegraphics{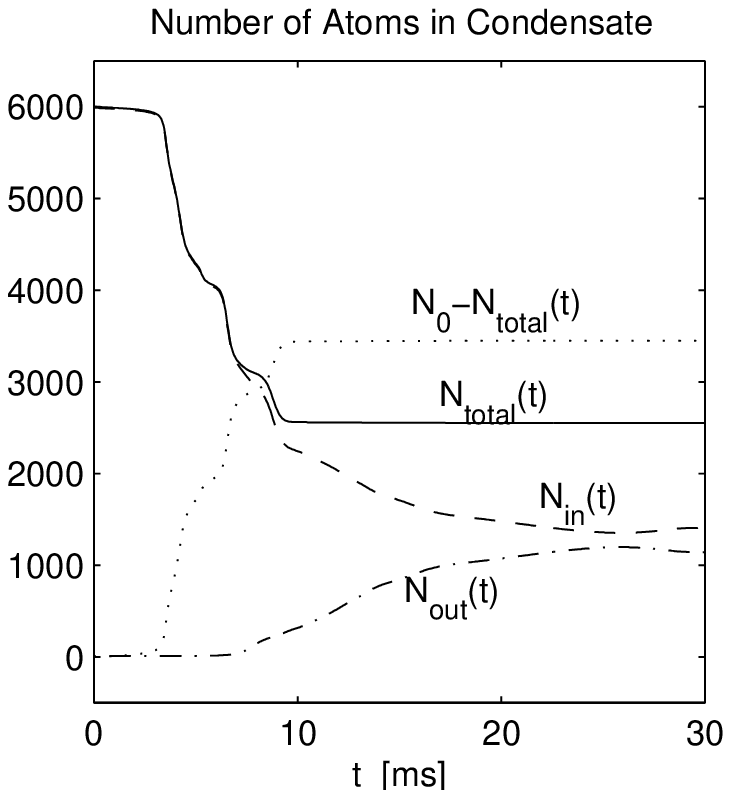} 
\caption{a) Radial and axial potential energy per particle outside 
the core of the condensate as functions of time 
$\tau_{\rm evolve}$ [ms] (labelled as $r$)
and b) number of atoms inside and 
outside of the core of the condensate. We have chosen $a_{\rm 
init}=0$, $N_0=6000$, and $a_{\rm collapse}=-30a_0$. 
\label{fig10}} 
\end{figure}

Choosing a core domain $\Omega_0=[-b_x,b_x]\times 
[-b_y,b_y]\times[-b_z,b_z]$ and domains $\Omega_z={\bf 
R}\setminus[-b_z,b_z]$, $\Omega_{xy} = {\bf R}^2\setminus [-b_x, 
b_x]\times [-b_y,b_y]$ \cite{footnote}, we calculate the 
expectation value of the axial and radial potential energy per 
particle  in the region $\Omega_z$ and $\Omega_{xy}$ respectively
\begin{eqnarray*} 
E_{\rm Axial}(t) &=&\frac{\int_{\Omega_z} \frac{1}{2} z^2 \phi_z(z,t)\;dz} 
{\int_{\Omega_z} \phi_z(z,t)\;dz},\\ 
E_{\rm Radial}(t) &=&\frac{\int_{\Omega_{xy}}\frac{1}{2} 
\left(\gamma_x^2 x^2+\gamma_y^2 y^2\right)\phi_{xy}(x,y,t)\;dxdy}{ 
\int_{\Omega_{xy}} \phi_{xy}(x,y,t)\;dxdy}, 
\end{eqnarray*} 
 and the number of atoms inside the core of the 
condensate 
\[N_{\rm in}(t)=N_0 \int_{\Omega_0} |\psi|^2\; d\bx, 
\qquad N_{\rm out}(t)= N_{\rm total}(t)-N_{\rm in}(t),\] where 
$b_x=b_y=1.5$ and $b_z=2.5$ from our simulation for $a_{\rm 
init}=0$. Figure \ref{fig10} shows these potential energies and 
the number of atoms in the condensate for $N_0=6000$ and $a_{\rm 
collapse}=-30a_0$. We find qualitative agreement between numerics 
and experiment in the number of burst atoms and the revival time, 
i.e.~the time when the radial energy $E_{\rm Axial}(t)$ has a 
minimum. However, the energy at which the atoms are emitted are 
far too small in our simulation and rather correspond to the 
energies at which the jet atoms are formed. Within our model of 
the three dimensional GPE with three particle loss we were only 
able to find particles emitted from the condensate at energies 
comparable to those observed experimentally if we decreased the 
three particle loss rate as done in the simulations of 
\cite{Ueda}. Decreasing the three particle loss rate leads to a 
more pronounced peak in the wave function and higher particle 
densities during the collapse and thus to higher kinetic energies 
of the particles leaving the condensate. However, such small 
values of the three particle loss rate are ruled out by the above 
considerations in Sec.~\ref{res1}. Therefore we conclude that the 
three dimensional GPE with three particle loss is not able to 
describe all the features observed in the experiment \cite{Donley} 
correctly and some additional physical mechanisms like those 
investigated in \cite{Javanainen, Burnett, Holland, Stoof} need to 
be taken into account. 
 
\section{Conclusions} 
\label{concl} 
 
In conclusion we have shown that the GPE describes the physics of 
collapsing and exploding condensates apart from the energies of 
the burst atoms for the chosen three particle loss rate $K_3^0$. 
We also found that no value for $K_3^0$ reproduces all the aspects 
of this experiment correctly. We obtained excellent agreement for 
the number of remnant atoms and the collapse times. Also the jets 
of atoms are well reproduced by the GPE and we found that small 
variations in the initial condition for the wave function yield 
big changes in the number of jet atoms. The large fluctuations in 
the number of jet atoms observed in the experiment could thus be 
due to uncertain initial conditions. 
 
\acknowledgments 
W.B. acknowledges support  by the National University of 
Singapore. P.A.M. acknowledges support from the EU-funded network 
'HYKE', and from his WITTGENSTEIN-AWARD, funded by the Austrian 
National Science Fund FWF. D.J acknowledges support from the 
WITTGENSTEIN-AWARD of P. Zoller. This research was supported in 
part by the International Erwin Schr\"{o}dinger Institute in 
Vienna.


\begin{thebibliography}{99}
\bibitem{footnote} 
We use domains defined in cartesian coordinates rather than 
adhering to the cylindrical symmetry of the problem as this 
corresponds more closely to the experimental way of measuring the 
corresponding properties.
\end{thebibliography}
\end{document}